\documentclass[aps,epsfig,twocolumn]{revtex4}

\usepackage{graphicx}
\usepackage{dcolumn}
\usepackage{bm}

\newcommand{\bq}{\begin{equation}}
\newcommand{\eq}{\end{equation}}
\newcommand{\bqa}{\begin{eqnarray}}
\newcommand{\eqa}{\end{eqnarray}}
\newcommand{\nn}{\nonumber \\}

\begin{document}
\draft
\title{ 
Vortex description of the fractionalized phase in exciton bose condensate
}

\author{Sung-Sik Lee$^{1}$, T. Senthil$^{1,2}$ and Patrick A. Lee$^{1}$}
\address{~$^{1}$ Department of Physics, Massachusetts Institute of Technology,\\
Cambridge, Massachusetts 02139, U.S.A.\\
~$^{2}$ Center for Condensed Matter Theory, Department of Physics,
Indian Institute of Science, Bangalore 560 012, India\\
}
\date{\today}
       
\begin{abstract}
As a sequel to the previous work 
[Phys. Rev. B {\bf 72}, 235104 (2005)]
we present a vortex description of the
fractionalized phase in exciton bose condensate.
Magnetic flux line and monopole of the 3+1D 
emergent U(1) gauge theory
are identified in the exciton picture.
A bundle of vortex/anti-vortex pairs of all flavors of excitons 
corresponds to the magnetic flux line and
a point at which the vortices and anti-vortices recombine is
identified as magnetic monopole.
This completes the magnetic sector of the low energy excitation
in the fractionalized phase.

\end{abstract}
\maketitle

\section{Introduction}
Recently, various fractionalized phases have been studied in 
exciton bose condensate\cite{LEE05}.
It was shown that a single exciton model can support
various fractionalized phases with either fractionalized 
boson or fermion along with photon.
The world line representation of the fractionalized particle 
and the emergent photon was constructed from the 
world lines of the exciton.
The confinement, Coulomb and Higgs phases were
described in terms of the dynamics of the web made of 
exciton world lines in a space-time picture. 
The world line representation 
turned out to be
most useful in describing 
the confinement/deconfinement phase
transition of the fractionalized particle 
which is electrically
charged with respect to the emergent photon. 
This is because the world line representation
corresponds to an electric representation 
of the emergent gauge theory
which keep tracks of the electric charge and flux line.
On the other hand, it was hard to describe magnetic 
charge and flux in the
electric representation.
This is attributed to the uncertainty
principle.
The electric degrees of freedom 
need to be condensed in order for
the magnetic excitations to be well defined\cite{DUAL}.
Thus it is desirable to use a magnetic description to 
understand magnetic excitations of the emergent gauge theory
in the exciton picture.
Recently, various 3+1D models have been proposed to show fractionalization
\cite{MOTRUNICH2,WEN,MOTRUNICH1,MOESSNER,HERMELE}.
In Ref. \cite{MOTRUNICH2}, fractionalized phase in a 3+1D bosonic model
has been studied in the magnetic description.
In this picture, the emergent photon is understood 
as a long wavelength fluctuation of
condensed vortices of multiple species.
The objective of the present paper is to employ
the vortex description\cite{MOTRUNICH2} to
study the fractionalized phase 
in the exciton bose condensate\cite{LEE05}.
We identify magnetic flux and monopole 
of the emergent gauge theory
in the exciton picture.

The fractionalized phase in the exciton system can also be 
understood based on conventional slave-particle approaches.
The slave-particle theory inevitably involves an infinitely 
strong gauge coupling because the gauge field comes from the constraint field.
The infinite bare gauge coupling ensures that the slave-particles
are confined at high energy.
Although the microscopic slave-particles are always confined within excitons,
the fractionalization can still occur 
as the gauge coupling is renormalized to a finite value at low energy.
The field theoretic argument for the gauge coupling renormalization
in the slave-particle theory is consistent with the
dual world line description of exciton\cite{LEE05}.
From this analysis, it is shown that the fractionalized phase occurs 
at least in the limit of $N$ sufficiently large 
where $N$ is the degeneracy of bands\cite{LEE05}.
However, the critical $N_c$, above which
fractionalization occurs for the exciton model, is not known. 
Numerical simulation is necessary to study this issue.
The objective of the present paper is not to address the 
occurrence of fractionalization for a given finite $N$.
Instead, we would like to understand magnetic excitations
of the emergent gauge theory in terms of the original
exciton in the fractionalized phase with large $N$.
The identification of magnetic degrees of freedom 
also provides an alternative way of probing emergent photon
in the fractionalized phase.
This will be useful in numerical simulation 
based on the number representation of exciton.
In this paper, $N$ is assumed large enough that the model has a fractionalized phase although
some examples are given with a small $N$ 
for the sake of illustration.

\begin{figure}
        \includegraphics[height=6cm,width=6cm]{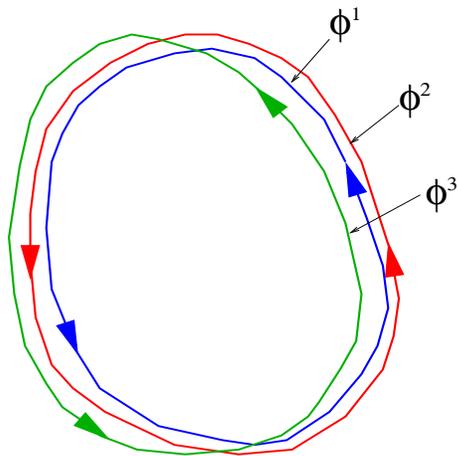}
\caption{
(color online)
Bundle of vortices in space for $N=3$.  
Each line represents vortex in the phase of $\phi^a$              
as indicated in the figure.
}
\label{fig:bundle}
\end{figure}

\begin{figure}
        \includegraphics[height=9cm,width=7cm]{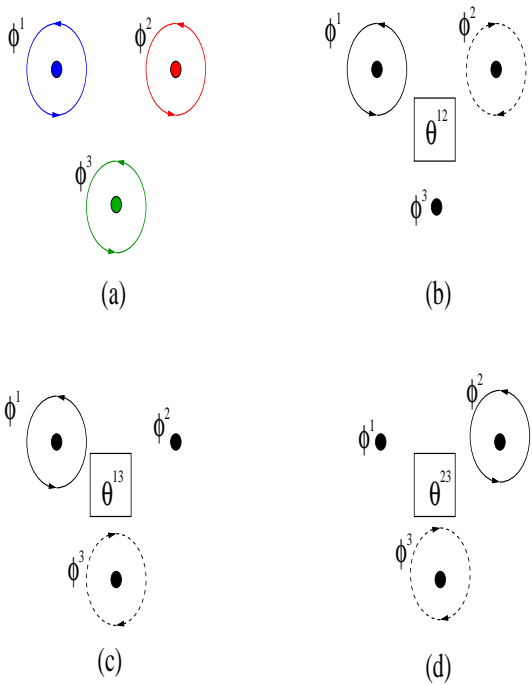}
\caption{
(color online)
A vortex bundle cut by an x-y plane with a fixed $\tau$ and $z$ for $N=3$.
(a) denotes three vortices which have vorticity $1$ for each $\phi^a$ with
$a=1,2,3$.
(b), (c) and (d) represent            
the configurations of exciton phases,           
$\theta^{12}$,
$\theta^{13}$,
and $\theta^{23}$ respectively,
in the presence of the vortex bundle.
The solid line denotes vorticity $+1$ and
the dashed line, $-1$.
}
\label{fig:bundle_section}
\end{figure}

Here is an overview of the present paper.
We focus on 3+1D and the parameter
regime where the exciton bose condensate is described 
by the off diagonal phase modes $\theta^{ab}$ of the 
$N \times N$ Hermitian matrix model,
\bqa   
S  & = &
- \frac{K}{4} \sum_{<i,j>} tr^{'} ( \chi_i^\dagger \chi_j + h.c.)
 + \sum_{n \geq 3} \tilde K_n \sum_i tr^{'} \chi_i^n.
\label{H}
\eqa
Here $\chi^{ab} = \chi_o e^{i \theta^{ab}}$ is 
the Hermitian matrix element 
of fixed amplitude $\chi_o$
with $\theta^{ab} = -\theta^{ba}$.
$\chi^{ab}$ describes
the bose condensate of the exciton
made of a particle in the $a$-th band and 
a hole in the $b$-th band where
$a,b = 1,...,N$ are band (flavor) indices. 
$K$ and $\tilde K_n$ are coupling constants determined 
from microscopic model.
With the strong third order interaction ($|\tilde K_3 \chi_o^3| >> 1$, $\tilde K_3 < 0$), 
the phase of the Hermitian matrix is further constrained
to satisfy 
\bq
\theta^{ab} = \phi^a - \phi^b,
\label{c}
\eq
where $\phi^a$ is the phase of slave boson.
The slave boson is introduced to
parameterize the low energy modes of exciton.
Details can be found in Ref. \cite{LEE05}.
Here we start with the Higgs phase 
where $\phi^a$ are coherent.
The elementary excitations in the Higgs phase
are super current modes and vortices.
There are $\frac{N(N-1)}{2}$ different vortices
of the off diagonal phase modes of the Hermitian matrix.
However only $N$ of them remain as low energy excitations
because of the dynamical constraints in Eq. (\ref{c}).
Each low energy vortex can be represented as vortex of 
the slave boson $\phi^a$ with $a=1,...,N$.
Note that vortex of a slave boson $\phi^a$
involves vortex (anti-vortex) of 
$\theta^{ab}$ ($\theta^{ba}$)
for all $b$ with $a<b$ ($a>b$).
Vortices of the slave bosons with different flavors
attract each other.
This is because two vortices of $\phi^a$ and $\phi^b$
involves a vortex (anti-vortex) of $\theta^{ab}$ 
at the position of the vortex of $\phi^a$ ($\phi^b$).
Because of the attraction, they
can form a bundle where vortices of all flavors 
are bound with
each other within a finite distance.
We will refer to this object as vortex bundle.
A vortex bundle in $N=3$ case is displayed in Fig. \ref{fig:bundle}.
The vortex bundle is special among vortex excitations
because it does not have net vorticity of the exciton phase.
In the exciton picture the vortex bundle corresponds to
pairs of vortex and anti-vortex 
as is shown in Fig. \ref{fig:bundle_section} for $N=3$.
Thus there is no long range interaction between
segments of the  vortex bundle. 
Vortex and anti-vortex of exciton 
can recombine at a point and
the vortex bundle can end at a point in space 
as is shown in Fig. \ref{fig:segment}.
The end point represents a point particle.
In determining physical property of the particle,
the orientedness of the vortex bundle is important.
The vortex bundle has orientation
even though a pair of vortex and anti-vortex of individual
exciton does not have orientation.
The orientation comes from correlation between 
vortex/anti-vortex pairs of different flavors.
For example, there are two possibilities depending on whether
the positions of vortices of $\theta^{ab}$  and $\theta^{ac}$
are coincident or the positions of anti-vortices are coincident.
If the positions of vortices (anti-vortices) coincide, 
it is represented as a vortex (anti-vortex) of $\phi^a$ 
at the coincident position, which determines the orientation 
of the vortex bundle.
Once the correlation in the vorticity of two exciton condensates 
is fixed, the rest are fixed by the 
constraint (\ref{c}).
Since there is an orientation in the vortex bundle,
one end is identified as a particle and 
the other end, as an anti-particle
which is distinct from the particle.
In space-time picture the end points form a closed loop
which represents vacuum fluctuations of
the particle/anti-particle
as is shown in Fig. \ref{fig:boundary}. 
This is the world line picture of 
a charged particle
connected by a U(1) gauge flux line.
We interpret the vortex bundle 
as a flux line of the emergent gauge theory and
the end point of the vortex bundle as a particle 
carrying charge. 

Deep in the Higgs phase, all vortex excitations are gapped.
Sizes of both the individual vortex and vortex bundle are small.
The particle/anti-particle (end points of vortex bundle) 
are confined by short segments of the flux line (vortex bundle).
As the phase stiffness of the exciton condensate decreases,
the size of vortex increases.
If the size of vortex bundle diverges 
the end point of vortex bundle emerges as deconfined particle.
It can be accomplished by either condensation 
of individual vortices or condensation of vortex bundle 
without condensation of individual vortices.
First, consider the former case.
This corresponds to the Coulomb phase.
The Goldstone modes of the exciton are gapped by the 
individual vortex condensation.
With a finite mass gap of the deconfined particle,
the fluctuations of the vortex bundle
give rise to a gapless photon\cite{MOTRUNICH2}. 
The emergence of photon can be understood in the
same manner as the emergence of photon in the
phase transition from the confinement to Coulomb phases\cite{LEE05}.
The difference is that the confinement/Coulomb phase
transition involves the divergences in the size 
of the world line web of exciton\cite{LEE05}
while the Higgs/Coulomb phase transition involves 
the world sheet of vortex bundle.
Since the world line web of exciton was identified as
the world sheet of electric flux line of the emergent gauge theory,
we identify the vortex bundle as magnetic flux. 
Accordingly the end point of the vortex bundle is identified 
as magnetic monopole.
This identification comes from the fact that the
Higgs phase is electro-magnetically dual to the confinement phase. 
Note that the term `electric' and `magnetic' has only relative meaning here.
Since we choose to call the particle (flux) which is confined
in the confinement as electric charge (flux), we call
the charge (flux) which is confined in the Higgs phase 
as magnetic charge (flux).
It is noted that the Coulomb phase occurs 
when vortices condense without 
the condensation of the magnetic monopoles.
If magnetic monopoles are condensed, the emergent photon
is gapped owing to the dual Higgs mechanism\cite{DUAL}.
One can understand that this Coulomb phase generically occurs 
in a large N limit where 
$N$ becomes large while 
the phase stiffness of the slave bosons is fixed.
(The phase stiffness of the original exciton should be scaled
as $1/N$ in order to fix the phase stiffness of slave bosons\cite{LEE05}.)
In this limit, the mass of the magnetic monopole is proportional
to $N$ because superfluidity of the $N$ slave bosons 
is suppressed near the position of the monopole.
In this large N limit, the monopole becomes very massive
and the monopole generically remains gapped when
the vortices of slave bosons condense.

Second, consider the case where only vortex bundle condenses
without the condensation of individual vortices.
The magnetic monopole is also deconfined and
the photon  emerges.
The difference from the first case is that
the Goldstone modes of the exciton remain gapless.
This is because the condensation of the vortex bundle alone
does not disorder the phase of exciton.
Thus we refer to this phase as Higgs$^*$ phase to distinguish it
from the usual Higgs phase which does not have emergent photon.
This is the analogy of AF$^*$ phase where fractionalization coexists 
with antiferromagnetic long range order\cite{AFstar}.
However, it is noted that the Higgs$^*$ is less likely to occur 
than the Coulomb phase.
This is because it needs condensation of vortex bundle while individual
vortices of slave bosons are `gapped'.
Since both the phase stiffness of the vortex bundle and the mass of the
magnetic monopole are proportional to N, the Higgs$^*$ is not guaranteed
to occur even in the large N limit.
One may need to introduce and fine tune other interactions in the microscopic model
in order to achieve the condensation of vortex bundle without the condensation
of magnetic monopoles. 
Finally, if monopole is condensed then electric charge is confined
because of the uncertainty principle.
This is the confinement phase.
The schematic phase diagram is shown in 
Fig. \ref{fig:phase} and Fig. \ref{fig:phase_2}.

\section{Vortex description of the fractionalized phase in the Hermitian matrix model of exciton bose condensate}

Since the constraint Eq. (\ref{c}) satisfies the potential energy in Eq. (\ref{H}),
only the kinetic energy remains and
the partition function of the Hermitian matrix of the exciton condensate 
can be written as a functional integral over the slave
bosons (see. Eq. (25) of Ref. \cite{LEE05}),
\bqa
Z & = &
\int_0^{2\pi} D \phi^a 
e^{
\frac{\kappa}{2} \sum_{i,\mu} \sum_{a<b} 
\cos \left( \nabla_\mu \phi^a(i) - \nabla_\mu \phi^b(i) \right)
}.
\label{z1}
\eqa
Here 
$\kappa = 2 K \chi_o^2$, 
$i$ is the site index in the 3+1D Euclidean lattice and
$\nabla_\mu \phi^a(i) = \phi^a(i + \mu ) - \phi^a(i)$
with $\mu$, the link direction.
A dual transformation similar to that of the 3+1D XY-model\cite{PESKIN,SAVIT,MOTRUNICH2}
leads to a vortex representation for the slave boson $e^{i \phi^a}$.
For details of the duality transformation see Appendix A.
In the vortex representation, the partition function becomes
\begin{widetext}
\bqa
Z & = &
\sum_{F_{\rho \sigma}^a}
\int D g_{\rho \sigma}^a
\exp \Bigl(
-\sum_I 
\Bigl[
\frac{1}{\kappa N}
\sum_{a,\mu}
\left( 
\frac{1}{4\pi} P_{ab} \epsilon_{\mu \nu \rho \sigma} \nabla_\nu g_{\rho \sigma}^b (I)
\right)^2 
+ \frac{i}{2} g_{\rho \sigma}^a(I) P_{ab} F_{\rho \sigma}^b(I)
\Bigr]
\Bigr)
\delta \left(   
\nabla_\sigma F_{\rho \sigma}^a(I)
\right). \nn
\eqa
\end{widetext}
Here we closely followed the notations of Ref. \cite{MOTRUNICH2}.
$I$ is the index of the dual lattice,
$\mu$ ,$\nu$, $\rho$, $\sigma$, the link direction, 
and $a$, $b$, the flavor indices.
Repeated indices are understood to be summed.
$F_{\rho \sigma}^a(I)$ is an integer which represents 
the presence of the world sheet of 
a vortex of $\phi^a$ at the $\rho \sigma$ plaquette
of site $I$.
$g_{\rho \sigma}^a(I)$ is the two-form dual gauge field
coupled to the world sheet of vortex.
It mediates interaction between vortices.
$P_{ab} = \delta_{ab} - \frac{1}{N}$ is a projection operator
in the flavor space.
Among the $N$ dual gauge fields,
there is one zero mode which has eigenvalue $0$ of $P$.
It is along the vector
${\bf v}_0^T = \frac{1}{\sqrt{N}}(1, 1, \cdots, 1)$.
Thus the flavor independent mode $g_{\rho \sigma}^a (I) =g_{\rho \sigma} (I)$
is projected out.
The zero mode corresponds to
an unphysical mode 
with $\phi^a = \phi$.
This mode is spurious because the flavor independent shift in every $\phi^a$ 
does not make any change in the phase of exciton $\theta^{ab} = \phi^a - \phi^b$.
The constraint $\nabla_\sigma F_{\rho \sigma}^a(I) = 0$
ensures that the world sheet of each vortex 
should be closed in the 3+1D space-time.
For notational simplicity, from now on we will omit the site index $I$ in the field variables.

\begin{figure}
        \includegraphics[height=5cm,width=8cm]{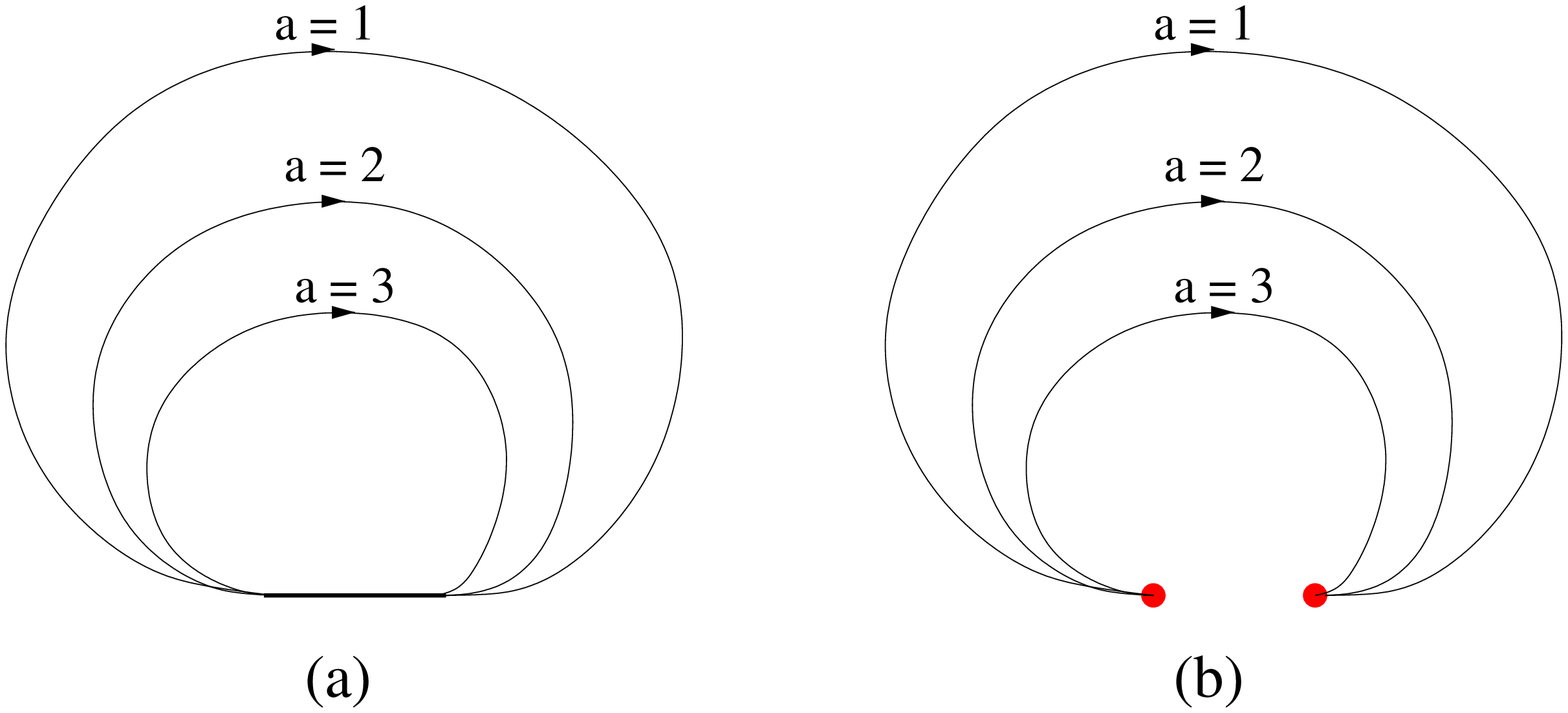}
\caption{
(color online)
A space picture of vortex bundle with overlapped segment for $N=3$.
The region where vortices of all flavors overlap in (a)
is the same as vacuum.
With the overlapped region removed in (b),
the vortex bundle end at two points which
are denoted as red (gray) dots in the figure.
}
\label{fig:segment}
\end{figure}

\begin{figure}
        \includegraphics[height=6cm,width=6cm]{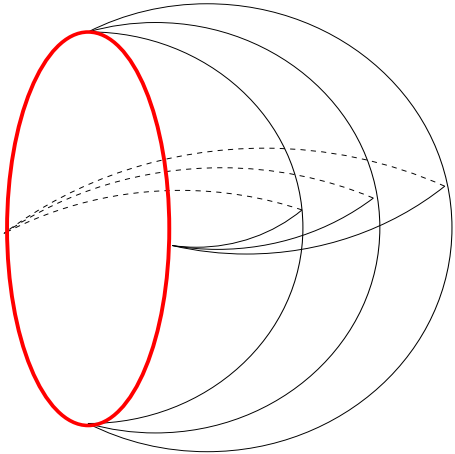}
\caption{
(color online)
Space-time picture of world sheets of vortices with a boundary for $N=3$.
The red (gray) thick line represents the boundary of the vortex bundle.
}
\label{fig:boundary}
\end{figure}

Because of the projection in the dual gauge field,
one mode of $F_{\rho \sigma}^a$ which is along
the ${\bf v}_0$ direction in the flavor space has no dynamics.
This mode represents the overlapped vortices in all $\phi^a$ 
on a same plaquette. 
It is also a spurious mode since the exciton phase is not distorted at all by the 
coincident $N$ vortices.
Thus the region where $N$ vortices are overlapped on a same plaquette
should be regarded as vacuum.
An example for the $N=3$ case is displayed in Fig. \ref{fig:segment}.
With the overlapped region removed, 
the world sheet of vortex is no longer closed.
Lines of $N$ vortices can end together at a point in space 
as is shown in Fig. \ref{fig:segment} (b).
In space-time picture, world sheets of the vortex bundle 
made of $N$ vortices can end 
on a closed loop as is shown in Fig. \ref{fig:boundary}.
We interpret the closed loop
as a world line of a charged `particle'.
The vortex bundle which emanates
from the world line of the particle is a `flux'.
In the exciton language, the flux corresponds to
a bundle of vortex/anti-vortex pairs of all off diagonal 
excitons $e^{i\theta^{ab}}$ as was explained in the introduction 
and shown in Fig. \ref{fig:bundle_section}. 
The charged particle is the end point where the
vortex and anti-vortex recombine.
Now, we discuss the flux and the particle
in the gauge theory picture.
The slave bosons $e^{i \phi^a}$ carry U(1)
gauge charge and is electrically coupled to a compact U(1)
gauge field\cite{LEE05}.
Vortices of the charged particle should involve flux of the coupled gauge field.
The energy cost is 
$\rho_s \sum_a ( \partial_\mu \phi^a - a_\mu )^2$.
The vortex bundle is accompanied by $2\pi$ 
magnetic flux of the gauge field
because it involves vortices of 
slave bosons of all flavors.
On the other hand, the energy of an individual
vortex which involves the winding of a single
$\phi^a$ is minimized by a $2\pi/N$ flux.
The end point of the vortex bundle is a magnetic monopole
which is the source of the $2 \pi$ magnetic flux line.
Taking into account the open boundary of the vortex world sheets
we rewrite the partition function as
\bqa
Z & = &
\sum_{F_{\rho \sigma}^a}
\sum_{l_{\rho}}
\int D g_{\rho \sigma}^a
D c_\sigma^a
D \alpha
e^{-S},
\eqa
where
\bqa
S & = &
\sum_I 
\Bigl[
\frac{1}{\kappa N}
\sum_{a,\mu}
\left( 
\frac{1}{4\pi} P_{ab} \epsilon_{\mu \nu \rho \sigma} \nabla_\nu g_{\rho \sigma}^b
\right)^2
+ \frac{i}{2} g_{\rho \sigma}^a P_{ab} F_{\rho \sigma}^b \nn
&& 
+ \frac{1}{2 \kappa_m} \sum_\rho l_\rho^2 
+ \frac{1}{2 \kappa_v} \sum_a \sum_{\rho < \sigma} (F_{\rho \sigma}^a)^2 \nn
&& 
- i c_\rho^a \left( \nabla_\sigma F_{\rho \sigma}^a - l_\rho \right)
+ i \alpha \left( \nabla_\rho l_\rho \right)
\Bigr].
\eqa
Here $l_\rho$ represents the world line of the magnetic monopole.
$1/\kappa_m$ is the mass of the monopole
and $1/\kappa_v$, the tension of the vortex world sheet.
The bare mass and tension is zero in the lattice scale, 
i.e., $1/\kappa_m = 1/\kappa_v = 0$.
However they are renormalized to nonzero values in 
long distance scale.
$c^a_\sigma$ and $\alpha$ are Lagrangian multipliers imposing
the flux conservation condition
$ \nabla_\sigma F_{\rho \sigma}^a = l_\rho$
and the current conservation of magnetic monopole
$ \nabla_\sigma l_\sigma = 0$ respectively.

Besides lattice scale $\zeta_1$, there are three length scales in this theory.
They are the size of the monopole world line $\zeta_2$,
the size of individual vortex $\zeta_3$,
and the size of the vortex bundle $\zeta_4$.
Note that it is possible that $\zeta_4 >> \zeta_3$. 
This is because $N$ vortices can form a vortex bundle 
whose effective tension is smaller than the tension of individual vortex.
In the Higgs phase, all of the length scales are finite.
With finite $\zeta_4$, the magnetic monopoles connected by the bundle of vortices are confined.
In the long wavelength limit, the vortex and the monopole/anti-monopole excitations 
can be ignored and
the low energy theory is described by the $N-1$ Goldstone modes $g_{\rho \sigma}^b$.
Note that there are only $N-1$ modes because of the projection $P_{ab}$.
As $\kappa$ decreases the tension (core energy) of the vortex decreases and
all of the $\zeta_2$, $\zeta_3$ and $\zeta_4$ increase.
There are different possibilities depending on which scale diverges.

If $\zeta_4$ diverges while $\zeta_2$ and $\zeta_3$ remain finite, 
the monopoles are deconfined without condensation of individual vortex.
Only the bundle of vortices condenses.
Since there is no long range interaction between the vortex bundle,
the low energy theory is described by gauge theory.
The emergence of photon is signified from the long range correlation
between the loop operators,
\bqa
\left< \delta T_{C_2}^\dagger \delta T_{C_1} \right>,
\eqa
where $\delta T_C = T_C - < T_C >$ and
$T_C$ is a creation operator of a vortex bundle 
along the loop $C$ (see Fig. \ref{fig:bundle}).
In the gauge theory picture it
creates magnetic flux line along $C$.
It is the 't Hooft operator\cite{THOOFT}
and is given by
$T_C = e^{i \oint_C \tilde a}$ where $\tilde a$ is the vector
potential associated with the dual field strength tensor,
$*f = d \tilde a$.
Here $(*f)_{\mu \nu} = \frac{1}{2}\epsilon_{\mu \nu \rho \sigma} f_{\rho \sigma}$
with $f_{\mu \nu} = \partial_\mu a_\nu - \partial_\nu a_\mu$,
the field strength tensor of the emergent gauge field.
Now we represent the 't Hooft operator in the exciton language.
In the Hamiltonian picture, we represent 
the 't Hooft loop operator as
\bqa
\hat T_C & \sim &  \sum_{|C_a - C| < L}
e^{
i \sum_{a<b} \sum_{\bf r}
\left( \varphi(C_a,{\bf r}) - \varphi(C_b,{\bf r}) \right)
\hat n^{ab}_{\bf r}
}.
\eqa
Here ${\bf r}$ is the 3 dimensional space vector.
$|C_1 - C_2|$ is a `distance' between two loops and
$L$, the maximum distance between vortex and anti-vortex
of exciton in a vortex bundle.
$\hat n^{ab}_{\bf r}$ is the number operator of exciton
of flavor $ab$ at site ${\bf r}$.
$\varphi(C,{\bf r})$ is given by
\bqa
{\bf \nabla} \varphi(C,{\bf r}) & = & \frac{1}{2} {\bf \nabla} 
\times \oint_C \frac{d {\bf r}^{'}}{ | {\bf r} - {\bf r}^{'} |}.
\eqa
$\varphi(C,{\bf r})$ changes by $2 \pi n_{C,C^{'}}$ 
when ${\bf r}$ moves along a loop $C^{'}$
which have linking number $n_{C,C^{'}}$ with the loop $C$.
Thus $\varphi(C,{\bf r})$ can have discontinuity by $2 \pi$ in space.
However the 't Hooft loop operator is well defined because  of
$e^{ i 2 \pi \hat n^{ab}_{\bf r} } = 1$.
To see that the above operator corresponds to the 't Hooft operator
we apply the operator to a phase eigenstate state of exciton. 
$\hat T_C$ create vortex (anti-vortex) along $C_a$ ($C_b$) for the
phase of $\theta^{ab}$ because
$e^{ i \varphi(C_a,{\bf r}) \hat n^{ab}_{\bf r} }$ 
($e^{ -i \varphi(C_b,{\bf r}) \hat n^{ab}_{\bf r} }$)
rotates $\theta^{ab}$ by 
$\varphi(C_a,{\bf r})$ 
($\varphi(C_b,{\bf r})$)
which, in turn, winds by 
$2\pi$ ($-2\pi$)
around the contour 
$C_a$ ($C_b$).
Thus $T_C$ creates vortex of slave boson $\phi^a$ 
along the contour $C_a$ for each $a$.
The vortex bundle is the bound state of these vortices
and the position of the individual vortices are summed 
around a given contour $C$ within a finite length scale $L$.
Since the vortex bundle corresponds to the magnetic flux line 
in the gauge theory picture, we identify $T_C$ as the
't Hooft operator of the gauge theory.
It is dual to the Wilson loop which was constructed in Ref. \cite{LEE05}.
The correlation function between electric (magnetic) flux is measured by $T_C$ ($W_C$).
In the Coulomb phase both of them have long range correlations.
With finite $\zeta_3$ the individual vortex is not condensed.
There still are long range correlation in the phases of exciton
and the $(N-1)$ Goldstone modes remain gapless.
This phase is a Higgs phase but it also has the emergent photon
as an additional gapless excitation apart from the Goldstone modes. 
Thus we call this phase as Higgs$^{*}$ to distinguish it from the 
conventional Higgs phase. 

If $\zeta_3$ diverges then the individual vortex is condensed
and the Goldstone modes are gapped.
This corresponds to the Coulomb phase.
In the Coulomb phase
it is convenient to use phase representation
for the vortex field rather than the world sheet of vortex. 
Note that $c^a_\sigma$ ($\alpha$) is conjugate to $F_{\rho \sigma}^a$ ($l_\rho$).
Thus $c^a_\sigma$ ($\alpha$) represents compact phase mode of vortex (monopole) field.
The summation over $F_{\rho \sigma}^a$ and $l_\sigma$ leads to the 
effective action for the phase variables,
\bqa
S & = &
\sum_I 
\Bigl[
\frac{1}{\kappa N}
\sum_{a,\mu}
\left( 
\frac{1}{4\pi} P_{ab} \epsilon_{\mu \nu \rho \sigma} \nabla_\nu g_{\rho \sigma}^b
\right)^2 \nn
&& 
- \kappa_v \sum_a \sum_{\rho < \sigma} 
\cos \left(
\nabla_\rho c_\sigma^a
- \nabla_\sigma c_\rho^a
- P_{ab} g^b_{\rho\sigma}
\right) \nn
&&
- \kappa_m \sum_\rho
\cos \left(
\nabla_\rho \alpha
-\sum_a c_\rho^a
\right)
\Bigr].
\eqa
In the Coulomb phase the vortex is condensed 
and $c^a_\sigma$ can be
regarded as non-compact variable.
In a rotated basis where the projection operator becomes diagonalized
as $P^{'}_{ab} = \delta_{ab} - \delta_{a1} \delta_{b1}$,
the Lagrangian becomes
\bqa
{\cal L} & = &
\frac{1}{\kappa N}
\sum_{a=2}^{N}
\sum_\mu
\left( 
\frac{1}{4\pi} \epsilon_{\mu \nu \rho \sigma} \nabla_\nu g_{\rho \sigma}^{a'}
\right)^2 \nn
&& 
+ \frac{\kappa_v}{2} 
\sum_{\rho < \sigma} 
\Bigl[
 ( \nabla_\rho c_\sigma^{1'}
- \nabla_\sigma c_\rho^{1'} )^2  \nn
&& + \sum_{a=2}^{N} 
 ( \nabla_\rho c_\sigma^{a'}
- \nabla_\sigma c_\rho^{a'} -  g^{a'}_{\rho\sigma})^2 
\Bigr] \nn
&&
- \kappa_m \sum_\rho
\cos \left(
\nabla_\rho \alpha
- \sqrt{N} c_\rho^{1'}
\right),
\eqa
where $c_\rho^{a'} = A_{ab} c_\rho^b$
with $A_{ab}$, the orthogonal matrix.
Especially, the first column vector
$A_{1a} = {1 \over \sqrt{N}}$ is the 
eigenvector for the zero mode of $P_{ab}$.
The non-zero modes $c_\rho^{a'}$  with $a \geq 2$
are coupled to the dual gauge field $g^{a'}_{\rho\sigma}$.
In the Coulomb phase (`Higgs' phase for the vortex field)
the non-zero modes can be absorbed into the longitudinal mode
of the dual gauge field
by the `gauge transformation',
\bqa
g^{a'}_{\rho\sigma} & \rightarrow &
g^{a'}_{\rho\sigma} + 
 ( \nabla_\rho c_\sigma^{a'} - \nabla_\sigma c_\rho^{a'} ).
\eqa
Then the dual gauge field acquire mass gap.
In the low energy limit, we obtain 
\bqa
{\cal L} & = &
 \frac{\kappa_v}{2N} 
\sum_{\rho < \sigma} 
 ( \nabla_\rho \tilde a_\sigma - \nabla_\sigma \tilde a_\rho )^2  \nn
&& - \kappa_m \sum_\rho
\cos \left( \nabla_\rho \alpha - \tilde a_\rho \right),
\eqa
where the photon field is given by
$\tilde a_\rho = \sqrt{N} c_\rho^{1'}$.
This is the low energy effective theory for the
monopole coupled to the non-compact U(1) gauge field.
Combined with the effective Lagrangian 
(Eq. (33) in Ref. \cite{LEE05})
which describes the coupling of the electrically charged 
particles (fractionalized bosons) to the gauge field $a_\mu$ 
(not $\tilde a_\rho$), they describe all of the low energy excitations 
in the Coulomb phase.
Note that the magnetic gauge coupling increases with increasing $N$
as $g_m^2 \sim \frac{N}{\kappa_v}$.
This is due to the Dirac quantization condition
$g g_m = \frac{1}{2}$, where $g$ is the electric gauge coupling.
The electric gauge coupling decreases with $N$ as
$g^2 \sim \frac{1}{N t^4}$\cite{LEE05} in the limit
where the effective hopping integral for the fractionalized boson
$t$ is fixed with increasing $N$.

\begin{figure}
        \includegraphics[height=7cm,width=8cm]{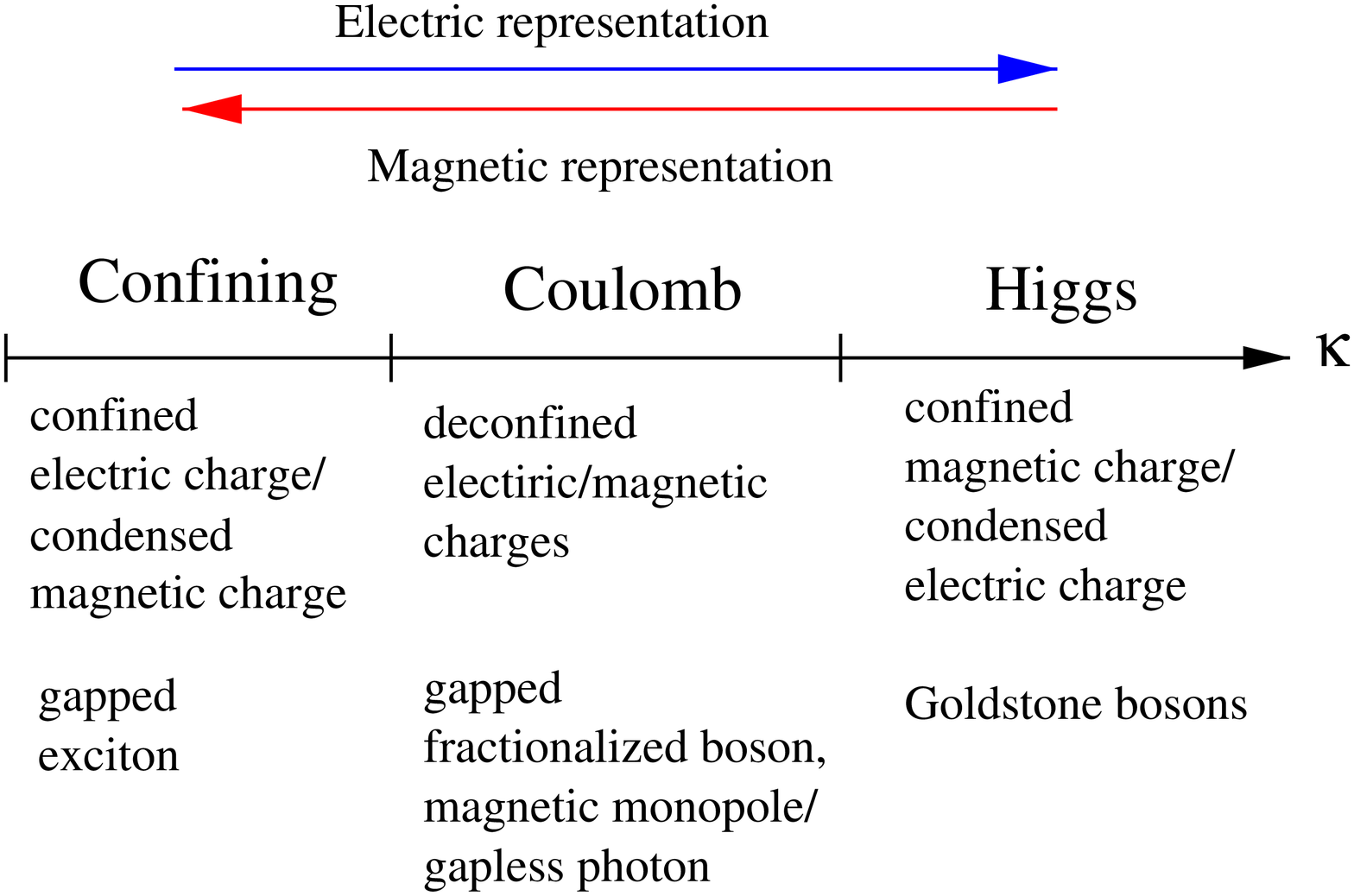}
\caption{
(color online)
Phase diagram of the Hermitian matrix model in the
strong coupling regime of the third order interaction.
The confined and condensed charges are shown along with
the low energy excitations in each phase.
The Higgs$^*$ phase is not shown in the diagram.
}
\label{fig:phase}
\end{figure}

\begin{figure}
        \includegraphics[height=2cm,width=8cm]{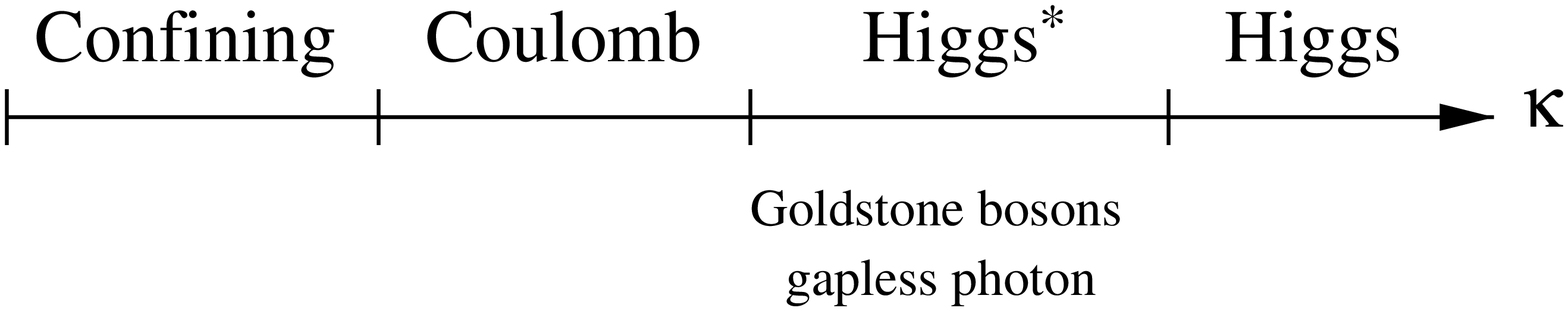}
\caption{
Alternative phase diagram.
The low energy particle of
Higgs$^*$ is the same as Higgs except
for the appearance of a gapless photon.
}
\label{fig:phase_2}
\end{figure}

Finally, if $\zeta_2$ diverges the monopoles are condensed.
With the monopole condensation the gauge field $\tilde a_\rho$
is gapped by Higgs mechanism.
Condensation of magnetic charge implies the confinement of
electric charge according to the uncertainty principle.
This corresponds to the confinement phase.
With this we complete the phase diagram of the Hermitian matrix model
from the Higgs phase side to the confinement phase side, while
the other direction was studied in Ref. \cite{LEE05}.
The schematic phase diagram is shown in Fig. \ref{fig:phase}.
Fig. \ref{fig:phase_2} shows an alternative phase diagram
which include the Higgs$^*$ phase.
The Higgs$^*$ and the Coulomb phases are exotic phases 
which have emergent photon.
However, we emphasize that whether those phases occur or not 
depend on the details of dynamics.
It is possible that either one or both of them are absent 
in the phase diagram of a specific model.

The low energy spectrum in the phase diagram of Fig. \ref{fig:phase}
is not quite symmetric despite the fact that the Higgs phase are dual
to the confinement phase.
This is attributed to the fact that there is only one kind of 
magnetically charged particle while there are $N$ electrically charged particles.
The condensation of electric charge leaves $N-1$ gapless modes in the Higgs phase
while there is no remaining gapless mode in the confinement phase.

\section{Conclusion}
In the present paper, we studied the fractionalized phase of exciton bose condensate
by using the vortex representation which is dual to the world line representation 
of exciton used in the previous work\cite{LEE05}.
From this we identified the magnetic flux and monopole excitations of the emergent gauge 
theory in terms of the exciton picture.
This completes the full identification of the low energy excitations in the fractionalized phase.

\newpage
\renewcommand{\theequation}{A\arabic{equation}}
\setcounter{equation}{0}
\section*{Appendix A. Vortex representation for the Hermitian matrix model}

We apply the vortex transformation to Eq. (\ref{z1}).
There are two equivalent ways of doing this.
In the first way, one represents the theory as a compact U(1) gauge 
theory coupled with $N$ slave bosons. 
The resulting action is 
\bqa
S^{'} & = & -t \sum_a \sum_{i,\mu} \cos( \nabla_\mu \phi^a(i) - a_\mu(i) ),
\eqa
where $t$ is the effective phase stiffness of the slave boson and
$a_\mu$, the compact U(1) gauge field.
Then the standard dual transformation followed by the integration of the 
gauge field leads to the vortex representation
as is shown in Eq. (\ref{app:vortex}).
In the second way, one can directly dualize the Eq. (\ref{z1}) 
without introducing gauge field in the intermediate step.
The two methods give rise to the same result.
Here we use the second method in order to 
emphasize the fact that the emergence of the U(1) gauge field
is not dependent of a particular way of introducing auxiliary field,
but is a consequence of the intrinsic dynamics of the model.

With the Villain approximation and the Hubbard-Stratonovich transformation,
the partition function (\ref{z1}) is rewritten as
\begin{widetext}
\bqa
Z & = &
\sum_{p^{a}_\mu(i)} \int_0^{2\pi} D \phi^a 
\int_{-\infty}^{\infty} D j_\mu^{ab}
\exp \Bigl( 
\sum_{i,\mu} \sum_{a<b} \Bigl[ 
 -\frac{1}{\kappa} (j_\mu^{ab}(i))^2 \nn
&& +i \left\{
\nabla_\mu \phi^a(i) - \nabla_\mu \phi^b(i) - 2 \pi ( p^a_\mu(i)  - p^b_\mu(i) )
\right\} j_\mu^{ab}(i)
\Bigr]
\Bigr). \nn
\eqa
\end{widetext}
Here $i$ is the site index in the 3+1D Euclidean lattice
and $\mu$, $\nu$ the link direction.
The integer field $p^a_\mu(i)$ is introduced in order to restore the periodicity
of the action as a function of $\phi^a$.
$p^a_\mu$ is decomposed into 
the divergenceless part and
the rotationless part as
$p^a_\mu = \tilde p^a_\mu + \nabla_\mu N^a$.
The summation over $N^a$ and the integration over $\phi^a$
leads to constraint
\bqa
\sum_b \nabla_\mu j^{(ab)}_\mu & = & 0,
\eqa
where
$j^{(ab)}_\mu \equiv j^{ab}_\mu$ for $a < b$,
$j^{(ab)}_\mu \equiv -j^{ba}_\mu$ for $a > b$ and
$j^{(aa)}_\mu \equiv 0 $.
Note that the current of each exciton is not conserved.
What is conserved is the flavor current
$\tilde j^a_\mu \equiv \sum_b j^{(ab)}_\mu$.
Introducing the flavor current we rewrite the partition function as
\bqa
Z & = &
\sum_{\tilde p^{a}_\mu} 
\int D j_\mu^{ab}
D \tilde j_\mu^{a}
D \lambda_\mu^{a}
\exp \Bigl(
- \sum_{i,\mu} \Bigl[
\sum_{a<b} \frac{1}{\kappa} (j_\mu^{ab})^2 \nn
&& + 2 \pi i \sum_a \tilde j_\mu^{a} \tilde p^{a}_\mu 
+ i \sum_a \lambda_\mu^a (
\tilde j_\mu^{a} -  \sum_b j^{(ab)}_\mu
)
\Bigr]
\Bigr)
\delta( \nabla_\mu \tilde j_\mu^{a} ), \nn
\eqa
where $\lambda_\mu^a$ is Lagrangian multiplier field
imposing the relation between the flavor current and 
exciton current.
Here we omit the site index $i$ in the field variables.
The Gaussian integration for $j^{ab}_\mu$ results in the partition function
\bqa
Z & = &
\sum_{\tilde p^{a}_\mu} 
\int D \tilde j_\mu^{a}
D \lambda_\mu^{a}
\exp \Bigl(
- \sum_{i,\mu} \Bigl[
\sum_{a,b} \frac{\kappa N}{4}
\lambda_\mu^a P_{ab} \lambda_\mu^b \nn
&& + i \sum_a \lambda_\mu^a \tilde j_\mu^a
+ 2 \pi i \sum_a \tilde j_\mu^{a} \tilde p^{a}_\mu
\Bigr]
\Bigr)
\delta( \nabla_\mu \tilde j_\mu^{a} ),
\eqa
where $P_{ab} = \delta_{ab} - {1 \over N}$.
$P$ has an eigenvalue $0$ and 
$(N-1)$-fold degenerate eigenvalue $1$.
The zero mode corresponding to the vanishing eigenvalue
is $\lambda^a = {1 \over \sqrt N}$.
This, in turn, project out the flavor independent
current $\tilde j^a_\mu = \tilde j_\mu$. 
This is the spurious mode as was discussed in the introduction.
The integration over $\lambda_\mu^a$ leads to the partition function
\bqa
Z & = &
\sum_{\tilde p^{a}_\mu} 
\int D \tilde j_\mu^{a}
\exp \Bigl(
- \sum_{i,\mu} \Bigl[
\sum_{a,b} \frac{1}{\kappa N}
\tilde j_\mu^a P_{ab} \tilde j_\mu^b \nn
&& + 2 \pi i \sum_a \tilde j_\mu^{a} P_{ab} \tilde p^{b}_\mu
\Bigr]
\Bigr)
\delta( \nabla_\mu \tilde j_\mu^{a} ).
\eqa
Solving the constraint $\nabla_\mu \tilde j_\mu^{a} = 0$ 
by introducing a two-form dual gauge field $g_{\rho \sigma}^a$,
$\tilde j_\mu^{a}  =  \frac{1}{4 \pi} \epsilon_{\mu \nu \rho \sigma} \nabla_\nu
g_{\rho \sigma}^a$,
and introducing a field $F_{\rho \sigma}^a$ describing the world
sheet of vortex,
$F_{\rho \sigma}^a = \epsilon_{\mu \nu \rho \sigma} \nabla_\mu \tilde p_\nu^a$,
we obtain the partition function
\bqa
Z & = &
\sum_{F_{\rho \sigma}^a}
\int D g_{\rho \sigma}^a
\exp \Bigl(
-\sum_I
\Bigr[
\frac{1}{\kappa N}
\sum_{a,\mu}
\left(
\frac{1}{4\pi} P_{ab} \epsilon_{\mu \nu \rho \sigma} \nabla_\nu g_{\rho \sigma}^b
\right)^2 \nn
&& + \frac{i}{2} g_{\rho \sigma}^a P_{ab} F_{\rho \sigma}^b
\Bigr]
\Bigr)
\delta \left(
\nabla_\sigma F_{\rho \sigma}^a
\right).
\label{app:vortex}
\eqa

\section{Acknowledgements}
The work of PAL and SSL was supported by the NSF grant DMR-0517222.
TS acknowledges funding from the NEC Corporation, the Alfred P. Sloan
Foundation, and an award from the The Research Corporation.

\end{document}